\documentclass[12pt]{article}
\usepackage{epsfig}

\def\beq{\begin{equation}}
\def\eeq#1{\label{#1}\end{equation}}
\def\eeqn{\end{equation}}

\def\beqa{\begin{eqnarray}}
\def\eeqa#1{\label{#1}\end{eqnarray}}
\def\eeqan{\end{eqnarray}}

\let\bar=\overbar

\def\Dslash{\not{\hbox{\kern-4pt $D$}}}
\def\dslash{\not{\hbox{\kern-2pt $\del$}}}

\def\BR{\mbox{\rm BR}}

\def\msb{{\bar{\ssstyle M \kern -1pt S}}}

\input{babarsym.tex}

\newcommand{\tautoel}  {\ensuremath{\tau \to e \nunub}\xspace}

\newcommand{\tautopi}  {\ensuremath{\tau \to \pi \nu}\xspace}

\newcommand{\tautoK}  {\ensuremath{\tau \to K \nu}\xspace}

\newcommand{\BRtauk}    {\ensuremath{\BR(\tautoK)}\xspace}
\newcommand{\BRtaupi}    {\ensuremath{\BR(\tautopi)}\xspace}

\newcommand{\tauknu}   {\ensuremath{ \tau^{-} \to K^{-} \nut}\xspace}
\newcommand{\BFtautoknu}    {\ensuremath{\BRtauk  }\xspace}
\newcommand{\BFtautopinu}    {\ensuremath{\BRtaupi}\xspace}

\newcommand{\gevccgevcc}{\ensuremath{{\mathrm{\,Ge\kern -0.1emV^2\!/}c^4}}\xspace}

\newcommand{\evcc}{\ensuremath{{\mathrm{\,e\kern -0.1emV\!/}c^2}}\xspace}

\newcommand{\vus}   {\ensuremath{|V_{us}|}\xspace}

\def\Title#1{\begin{center} {\Large {\bf #1} } \end{center}}

\begin{document}

\Title{Determination of $|V_{us}|$ from $\tau$ Decays }

\bigskip\bigskip

%+\addtocontents{toc}{{\it D. Reggiano}}
%+\label{ReggianoStart}

\begin{raggedright}  

{\it Ian. M. Nugent (representing the \babar
    Collaboration)\index{Nugent, I. M.}\\
III. Physikalisches Institut\\
 Physikzentrum\\
 RWTH Aachen\\
 52056 Aachen, Germany\\
Email: inugent@uvic.ca\\
Proceedings of CKM 2012, the 7th International Workshop on the CKM\\
Unitarity Triangle, University of Cincinnati, USA, 28 September - 2 October 2012
}
\bigskip\bigskip
\end{raggedright}

{\abstract The weak interaction between the first and second generation
of quarks, 
the Cabibbo-Kobayashi-Maskawa matrix (CKM) element \vus,  can be
probed using hadronic $\tau$ decays.  In this paper, we present the
recent measurements of hadronic $\tau$  decays from BELLE and \babar\ 
and the improvements in the determination of \vus from $\tau$
decays.
}

\section{Introduction}

Hadronic $\tau$ decays provide an opportunity to probe the relation of the first row of the Cabibbo-Kobayashi-
Maskawa (CKM) matrix by measuring the
coupling of the first and second generation of quarks to the weak
charged current, $|V_{us}|$~\cite{Cabibbo:1963yz}.  
Measurements of $|V_{us}|$ from $\tau$ decays are complimentary
to the kaon decay measurements\cite{Antonelli:2010yf}. The kaon measurements are 
consistent with the unitarity condition ($|V_{ud}|^2+|V_{us}|^2+|V_{ub}|^2 =1$), where
the value of $|V_{ud}|$ used in this comparison is provided from nuclear beta decays \cite{Hardy:2008gy}
and the contribution from $|V_{ub}|$ is
negligible\cite{Amhis:2012bh}. However, new
physics scenarios that couple primarily to the third generation could
cause deviation between measurements of $V_{us}$ in the kaon and
$\tau$ systems\cite{Krawczyk:2004na,Dorsner:2009cu,Loinaz:2002ep,Czarnecki:2004cw,Marciano:2007zz}. 

In $\tau$ decays, there are multiple techniques that can be used to
extract \vus. In this paper, we will limit ourselves to the three most
precise methods.   The technique that offers the potential for the
most precise measurement\cite{Maltman:2008ib} comes from the flavor breaking difference
with Finite Energy Sum 
Rules (FESR). More specifically,
\begin{eqnarray*}
\frac{R_{\tau,strange}}{|V_{us}|^{2}}-\frac{R_{\tau,non-strange}}{|V_{ud}|^{2}}=\delta
  R_{\tau,SU3\  breaking}
\end{eqnarray*}
where $R_{\tau,strange}=\Gamma(\tau^{-}\rightarrow X_{strange}
\nu_{\tau})/\Gamma(\tautoel)$ is the strange hadronic width,
$R_{\tau,non-strange}=\Gamma(\tau^{-}\rightarrow
X_{non-strange}\nu_{\tau})/\Gamma(\tautoel)$ is the non-strange hadronic
width
and $\delta R_{\tau,SU3\  breaking}$ is the theoretical SU(3) flavor
breaking
correction determined using Operator Product Expansion (OPE). 
From an experimental perspective, this technique requires that the
inclusive strange
and non-strange spectral density functions, which are constructed from
the sum of invariant mass distributions for each of the strange
and non-strange decay modes and
normalized to the corresponding branching fractions, are
measured. Since there are no solid predictions for the branching fractions
of hadronic individual $\tau$ decays, all possible modes must be
measured or have an upper bound placed on them. This technique is
completely 
independent of
the kaon measurements. If all of the branching fractions and
spectral functions are updated with the data from the
BELLE and \babar, this method
would be expected to make the most precise measurement of
\vus~\cite{Maltman:2008ib}.

Currently, the most precise technique for determining \vus from tau
decays is:
\begin{eqnarray*}
\frac{\BFtautoknu}{\BFtautopinu} = \frac{f_K^2 |V_{us}|^2}{f_\pi^2
  |V_{ud}|^2}
 \frac{\left( 1 -  \frac{m_K^2}{m_\tau^2} \right)^2}{\left( 1 -
 \frac{m_\pi^2}{m_\tau^2} \right)^2} (1+\delta_{LD}),
\end{eqnarray*}
where $f_K/f_\pi = 1.1936\pm  0.0053$~\cite{Laiho:2009eu} is
determined
from Lattice QCD, \Vud~\cite{Hardy:2008gy}, and
the long-distance correction $\delta_{LD} = (0.03\pm 0.44)\%$
is estimated~\cite{Banerjee:2008hg}
using corrections to $\tau\to h\nu_{\tau}$ and $h \to
\mu\nu_{\mu}$~\cite{Marciano:1993sh,Marciano:2004uf}. This method is
analogous to measurements in the kaon system and is sensitive to the
same Lattice QCD uncertainties. 

Measurements using the absolute branching fraction \tauknu,
\begin{eqnarray*}
BR(\tauknu) = \frac{G^2_F f^2_K \Vus^2 m^3_{\tau}
  \tau_{\tau}}{16\pi\hbar} \left (1 - \frac{m_K^2}{m_\tau^2} \right
  )^2 S_{EW},
\end{eqnarray*}
provides a competitive measurement compared to the former techniques,
however, it is also sensitive to Lattice QCD uncertainties. For this
method, the kaon decay 
constant is $f_{K}=156.1 \pm 1.1  MeV $~\cite{Laiho:2009eu}
 and the electroweak correction is $S_{EW}=1.0201\pm  0.0003 $~\cite{Erler:2002mv}.

\section{Experimental Results}

The B-Factories, \babar\ and BELLE, have measured many of the branching
fractions for the hadronic $\tau$ decay modes\cite{Amhis:2012bh}.
This includes the majority of the main
strange $\tau$ branching fractions, which are presented in Table \ref{table2},
as well as recent limits on unmeasured decay
modes\cite{Lees:2012de,Lees:2012ks}. This is in contrast to the small
number of measured invariant mass
spectra\cite{Lee:2010tc,SRyu:2012b,IMNugent:2012}.

%%%%%%%%%%%%%%%%%%%%%%%%%%%%%%%%%%%%%%%%%%%%%%%%%%%%%%%%%%%%%%%%%%%%%%%%%
%%
%%   use this format to include a LaTeX table  into your paper
%%
\begin{table}
\caption{The current status of the branching fraction for the strange
  $\tau$ decays.}
\footnotesize
\begin{center}
\footnotesize
\begin{tabular}{lcrr} 
\hline
\hline
Decay Mode & Branching Fraction (\%) & BELLE & \babar\ \\
\hline
\BFtautoknu        & $0.6955\pm0.096$ & &
\cite{Aubert:2009qj} \\
\BR($\tau^{-}\rightarrow K^{-}\pi^{0}\nu_{\tau}$) & $0.4322\pm0.0149$ &
&
\cite{Aubert:2007jh} \\
\BR($\tau^{-}\rightarrow K^{-}\pi^{0}\pi^{0}\nu_{\tau}\ (ex.\ K^{0})$)
&
$0.0630\pm0.0222$ &  &  \\
\BR($\tau^{-}\rightarrow K^{-}\pi^{0}\pi^{0}\pi^{0}\nu_{\tau}\ (ex.\                                                                                          
K^{0},\eta)$) & $0.0419\pm0.0218$ & &  \\
\BR($\tau^{-}\rightarrow K^{0}\pi^{-}\nu_{\tau}$) & $0.831\pm0.018$ &
\cite{Epifanov:2007rf} & \cite{Aubert:2008an} \\
\BR($\tau^{-}\rightarrow K^{0}\pi^{-}\pi^{0}\nu_{\tau}$) &
$0.3649\pm0.0108$
& \cite{SRyu:2012}& \cite{Paramesvaran:2009ec} \\
\BR($\tau^{-}\rightarrow K^{0}\pi^{-}\pi^{0}\pi^{0}\nu_{\tau}$) &
$0.0269\pm0.0230$ & & \\
\BR($\tau^{-}\rightarrow K^{0}h^{-}h^{-}h^{+}\nu_{\tau}$) &
$0.0222\pm0.0202$
& &  \\
\BR($\tau^{-}\rightarrow K^{-}\pi^{-}\pi^{+}\nu_{\tau}\ (ex.\ K^{0})$)
&
$0.2923\pm0.0068$ &\cite{Lee:2010tc} &\cite{Aubert:2007mh} \\
\BR($\tau^{-}\rightarrow K^{-}\pi^{-}\pi^{+}\pi^{0}\nu_{\tau}\ (ex.\                                                                                          
K^{0},\eta)$) & $0.0411\pm0.0143$ & &  \\
\BR($\tau^{-}\rightarrow K^{-}\eta\nu_{\tau}$) & $0.0153\pm0.0008$ &
\cite{Inami:2008ar} & \cite{delAmoSanchez:2010pc}\\
\BR($\tau^{-}\rightarrow K^{-}\eta\pi^{0}\nu_{\tau}$) &
$0.0048\pm0.0012$ &
\cite{Inami:2008ar} & \\
\BR($\tau^{-}\rightarrow K^{0}\eta\pi^{-}\nu_{\tau}$) &
$0.0094\pm0.0015$ &
\cite{Inami:2008ar} & \\
\BR($\tau^{-}\rightarrow K^{-}\omega\nu_{\tau}$) &
$0.0410\pm0.0092$ &
& \\
\BR($\tau^{-}\rightarrow K^{-}\phi\nu_{\tau} (\phi\to K^{-}K^{+})$) &
$0.0037\pm0.0014$ &
\cite{Lee:2010tc} & \cite{Aubert:2007mh} \\
 \hline
Total & $2.87(46)\pm0.04(98)$ \\
\hline
\multicolumn{4}{l}{Branching Fractions from HFAG 
  fit~\cite{Amhis:2012bh}
  $\chi^{2}$/d.o.f.=143.5/118 CL=5.5\% }\\
\hline
\hline
\end{tabular}
\label{table2}
\end{center}
\normalsize
\end{table}

\section{Discussion and Conclusion}

\begin{figure}[htb]
\begin{center}
\epsfig{file=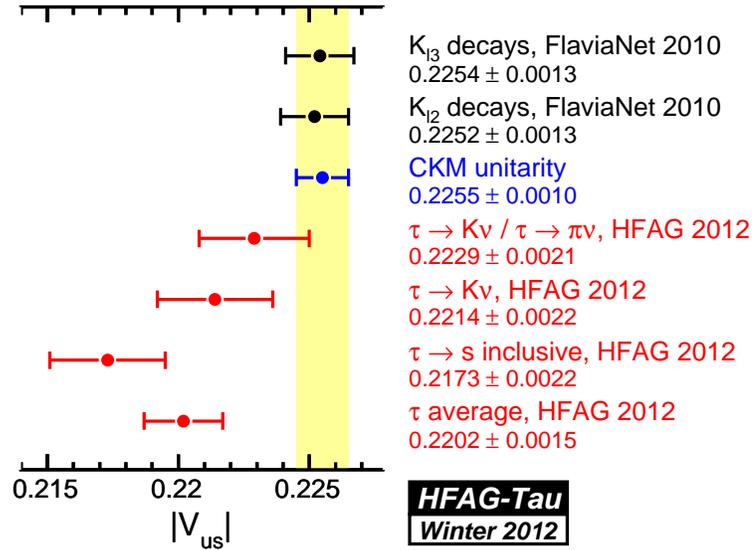,height=3.0in}
\caption{An update of \vus from the HFAG 2012 report\cite{Amhis:2012bh} for
  the hadronic $\tau$ decays. The HFAG values of \vus are extracted
  using the average branching fractions from HFAG. The three upper
  values are from $K_{l3}$ decays~\cite{Antonelli:2010yf}, $K_{l2}$ decays~\cite{Antonelli:2010yf} and
  the unitarity constraint~\cite{Hardy:2008gy}. }
\label{fig:vus}
\end{center}
\end{figure}

The HFAG value of $|V_{us}|$ extracted using the three techniques
mentioned above are compared to the
kaon measurements in Figure \ref{fig:vus}. In all of these methods, 
the uncertainty is limited
by the experimental precision. Both of the measurements of $V_{us}$ extracted from ratio of   
$\frac{\BRtauk}{\BRtaupi}$ and directly from \BRtauk are reasonably
consistent with unitarity determined from\cite{Hardy:2008gy}. Both of these
measurements are dominated by the \babar\ measurement \cite{Aubert:2009qj}.
The value of \vus extracted using the FESR method deviation from
unitarity is $3.4\sigma$. With the 
recent upper-limits on the unmeasured $\tau$ decay modes, the
possibility of this deviation resulting from missing decay modes is
becoming smaller. However, measurements of the hadronic $\tau$ decays
at BELLE and \babar\ seem to be systematically lower then the previous
world averages. This problem could be a result of differences in  
the definitions of the decay modes between the B-Factories and
previous experiments or an artifact from only having updated a subset
of all the $\tau$ hadronic branching fractions. 
Therefore, further results are needed before drawing any significant
conclusions.
On the theoretical side, the deviation could be related to convergence problems
with the weights employed for the FESR which are not taken in to
account by the systematic uncertainties~\cite{Maltman:2008ib,Kambor:2000dj,Maltman:2010hb}. 
 
%%%%%%%%%%%%%%%%%%%%%%%%%%%%%%%%%%%%%%%%%%%%%%%%%%%%%%%%%%%%%%%%%%%%%%%%%
%%
%%   use this format to include an .eps figure into your paper
%%
%%%%%%%%%%%%%%%%%%%%%%%%%%%%%%%%%%%%%%%%%%%%%%%%%%%%%%%%%%%%%%%%%%%%%%%%%%%

%\def\Discussion{
%\setlength{\parskip}{0.3cm}\setlength{\parindent}{0.0cm}
%     \bigskip\bigskip      {\Large {\bf Discussion}} \bigskip}
%\def\speaker#1{{\bf #1:}\ }
%\def\endDiscussion{}
%
%\endDiscussion
 
\end{document}